\documentclass[pre,twocolumn,superscriptaddress]{revtex4}
\usepackage{graphicx,epsfig}
\usepackage{epstopdf}

\DeclareGraphicsExtensions{. jpg,. pdf, . mps, .png, .eps, . ps, . EPS}
\usepackage{amsmath}
\usepackage{color}
\usepackage{afterpage}
\usepackage[section]{placeins}

\begin{document}
\newcommand{\beq}{\begin{equation}}
\newcommand{\eeq}{\end{equation}}
\newcommand{\bea}{\begin{eqnarray}}
\newcommand{\eea}{\end{eqnarray}}
\newcommand{\gt}{\tilde{g}}
\newcommand{\mt}{\tilde{\mu}}
\newcommand{\et}{\tilde{\varepsilon}}
\newcommand{\ct}{\tilde{C}}
\newcommand{\bt}{\tilde{\beta}}
\newcommand{\avg}[1]{\langle{#1}\rangle}
\newcommand{\Avg}[1]{\left\langle{#1}\right\rangle}
\newcommand{\cor}[1]{\textcolor{red}{#1}}

\title{Multilink Communities of Multiplex Networks}

\author{ Ra\'ul J. Mondrag\'on}
\address{ School of Electronic Engineering and Computer Science, Queen Mary University of London E1 4NS, United Kingdom}
\author{Jacopo Iacovacci}
\address{School of Mathematical Sciences, Queen Mary University of London E1 4NS, United Kingdom}
\author{Ginestra Bianconi}
\address{School of Mathematical Sciences, Queen Mary University of London E1 4NS, United Kingdom}

\begin{abstract}
Multiplex networks describe a large number of  complex social, biological and transportation  networks where a set of nodes is connected by links of different nature and connotation.  Here we uncover the rich community structure of multiplex networks by  associating a community to each multilink where the multilinks characterize the connections existing between any two nodes of the multiplex network. Our community detection method reveals the rich interplay between the mesoscale structure of the multiplex networks and their multiplexity. For instance some nodes can belong to many layers and few communities while others can belong to few layers but many communities. Moreover the multilink communities  can be formed by a different number of relevant layers. These results  point out that  mesoscopically there can be large differences in the compressibility   of multiplex networks. 
\end{abstract}

\maketitle
The current Big Data explosion requires the development of new algorithms and theoretical  methods  to extract information from large datasets. Often in this context, it is advantageous  to combine information coming from different sources and to represent the data by a  multiplex network \cite{boccaletti2014structure,kivela2014multilayer,bianconi2013statistical,buldyrev2010catastrophic,gomez2013diffusion}.  A multiplex  network is formed by a set of nodes connected in different layers by links indicating interactions of different types.   Multiplex networks are  ubiquitous spanning from  complex infrastructure networks~\cite{buldyrev2010catastrophic,shao2011cascade,huang2011robustness}, to   social ~\cite{nicosia2015measuring,mucha2010community,szell2010multirelational,menichetti2014weighted} biological ~\cite{nicosia2015measuring,bentley2016multilayer} and transportation networks ~\cite{de2014navigability,cardillo2013emergence}. For instance, individuals can be related by different type of social ties, neurons can interact through  chemical synapses and electrical gap junctions,  and  two locations can be connected by different means of transportation.

 A multiplex network tends to have a richer structure than single networks and this richness is reflected in its communities~\cite{mucha2010community,szell2010multirelational,battiston2016emergence,iacovacci2015mesoscopic,kao2017layer}.
The communities of a multiplex network cannot be obtained by considering its layers  individually. Some communities might exist only in one layer, other communities can overlap on many layers and finally there are  communities that only exist when considering the whole structure of the multiplex network. 
Several algorithms~\cite{mucha2010community,de2015identifying,jeub2017local,bennett2015detection,kuncheva2015community,lancichinetti2012consensus} have been recently proposed to detect multilayer communities. These include methods based on multilayer modularity optimization ~\cite{mucha2010community,bennett2015detection}, diffusion properties on multilayer networks ~\cite{de2015identifying,kuncheva2015community} and consensus clustering ~\cite{lancichinetti2012consensus}.
All these techniques are node-based community detection methods where each node or each replica-node (realization of a node in a given layer) is classified in one community.  Interestingly in the framework of single-layer community detection \cite{fortunato2016community,schaub2016many} it has been observed that link-based community detection methods~\cite{ahn2010link,evans2009line} can be very fruitful to describe the mesoscale organization of networks when nodes  belong to several communities at the same time \cite{palla2005uncovering}. The need to extend the link communities to multiplex network is rather pressing. For instance if we consider individuals interacting through different on-line social network platforms, say Twitter and Facebook, it might be misleading to think that an individual or an account (a Twitter or Facebook account) might belong just to a single community. In fact, influential Twitter of Facebook accounts tend to reach more than one community of the same online platform.

 In simple networks any two nodes can be either connected or not connected by a link, in multiplex network any two nodes can be connected in multiple ways. 
We say that two nodes are connected via a {\it multilink} \cite{bianconi2013statistical,menichetti2014weighted}, where the multilink describes the pattern of connections between two nodes.
In this work we propose a multilink community detection method for multiplex networks which extends link communities to  the multiplex network framework. Our community detection method is based on the similarity of incident multilinks. In order to reduce unnecessary layer-information, the similarity between two multilinks is measured by comparing the local structure of the multiplex against a { local, maximum entropy null model}. To avoid introducing bias via the null model, the null model describes our state of knowledge of the multiplex in a way that is maximally noncommittal to the layered structure.

Here we  show that using the proposed multilink community detection method not only we are able to extract relevant information on the mesoscale structure of multiplex networks, but also we can contribute to the scientific debate about the compressibility of the multiplex network structures.
Recent research on multiplex networks questions whether it is opportune to aggregate or disaggregate their layers. Aggregation of layers could be useful for removing redundant information.
De~Domenico~et~al.~\cite{de2015structural},  have shown that for the vast majority of multiplex networks there is trade-off between the information content and the minimization of their total number of layers.
The case of disaggregating a single network to a multi-layer network has been considered by Vales-Catala~et~al.~in Ref. \cite{valles2016multilayer}. According to their results some single networks are better  represented  as  multiplex networks because  they  are effectively the result of a blind multiplex network aggregation procedure.  Finally, Peixoto~\cite{peixoto2015inferring}, using a statistical inference approach, has revealed that there is no clear answer, the benefits of the aggregation or disaggregation of the layers are dependent on the system under study.

{Here we show that actually the optimal answer to the question whether it is more appropriate to aggregate or disaggregate a general multiplex network might not be global but mesoscale.} Our analysis of social, biological and transportation networks reveals that in  multiplex networks there is a very rich interplay between  their mesoscale organization and their multiplexity.
Multilinks communities can include connections of only one layer or of multiple layers. Additionally we observe that  not always the layer activity (in how many layers a node is connected) correlates with the community activity (in how many communities a node can be found). For example there can be nodes that are connected in many layers (high layer activity) but belong only to few multilink communities (low community activity) and nodes belonging to few layers (low layer activity) but belonging to many multilink communities (high community activity).
The first possibility suggests that mesoscopically the network could be compressed while the second possibility suggests that mesoscopically  the network could be expanded into many layers making a case for a definition of a mesoscale compressibility of the multiplex network.

\section*{RESULTS}

\subsection*{Multiplex network} Let us consider a multiplex network formed by $N$ nodes and $M$ layers $\alpha=1,2,\ldots, M$. The multiplex network is the set of $M$ networks $\vec{G}=\{G^{[1]},G^{[2]},\ldots, G^{[\alpha]},\ldots, G^{[M]}\}$ where each network $G^{[\alpha]}=\{V,E^{[\alpha]} \}$ is formed by the same set of $N$ nodes $V=\{i; i=1,2,\ldots, N\}$ and by the set of links $ E^{[\alpha]}$ which describe the connections in layer $\alpha$. We assume that all these networks are undirected and we represent each layer $\alpha=1,2,\ldots, M$ by the adjacency matrix ${\bf a}^{[\alpha]}$.
The  whole multiplex network can be expressed via its multilinks~\cite{bianconi2013statistical,menichetti2014weighted}.
Every pair of nodes $(i, j)$ is connected by a multilink 
\bea
\vec{m}_{ij}=\left(m_{ij}^{[1]},m_{ij}^{[2]},\ldots,m^{[\alpha]}_{ij}\ldots m_{ij}^{[M]}\right),
\eea
with $m_{ij}^{[\alpha]}=a_{ij}^{[\alpha]}$ indicating in which layers of the multiplex network the two nodes are connected. Whenever node $i$ and node $j$ are connected at least in one layer, i.e. $\vec{m} \neq\vec{0},$ we say that they are connected by a non-trivial multilink.
To decide if a non-trivial multilink exist, it is convenient to construct the aggregated network $\hat{G}$ formed by the $N$ nodes of the multiplex. The adjacency matrix ${\bf A}$ of the aggregated network $\hat{G}$ has elements 
\bea
{A}_{ij}=\theta\left(\sum_{\alpha=1}^M a_{ij}^{[\alpha]}\right),
\eea
where $\theta(x)$ is the step function $\theta(x)=1$ if $x>0$ and $\theta(x)=0$ if $x\leq 0$. We indicate with $L=\sum_{i<j}A_{ij}$ the total number of links of the aggregated network, or equivalently the number of non-trivial multilinks.

In a multiplex network the nodes might not be connected in each layer. The number of layers in which a node is connected (or active) is called the {\em node activity} \cite{nicosia2015measuring,cellai2016multiplex} and reveals relevant coarse grained information about the node.

\subsection*{Multilink similarity} In the context of single networks several community detection methods use hierarchical clustering applied either to a  similarity matrix  between nodes  \cite{ravasz2003hierarchical} or between links \cite{ahn2010link,evans2009line}. Here we construct a hierarchical clustering of multiplex networks based on a measure of similarity between incident multilinks.  
By defining the similarity between multilinks here we generalize  the  link communities previously defined  for single layers~\cite{evans2009line,ahn2010link} to  multiplex networks.

In a  similar spirit to the use of the  modularity function for detecting node communities \cite{newman2006modularity},  the similarity between  incident multilinks is evaluated by comparing simultaneously the cohesiveness and the multiplexity of their neighbourhood  to a  maximum entropy null model.

To every pair of  multilinks connecting nodes $i$ and $k$ and nodes $j$ and $s$ we assign the similarity $S_{ik,js}$. The  similarity  $S_{ik,js}$ is  non-zero only  between incident multilinks (i.e. for $s=k$) and is a function of two parameters: $\epsilon$ and $z$. The parameter $\epsilon\in (0,1)$ can be tuned depending on the role that we want to assign to the composition of the two incident multilinks with respect to their local neighborhood. The additional parameter 
$z\in (0,1)$ evaluates the role of multiplexity and represent the cost we want to attribute to incident multilinks of different composition.

Specifically  the non-zero similarities $S_{ik,jk}$ are given by
\bea
S_{ik,jk}=\epsilon \sigma_{ijk}+(1-\epsilon)\sigma_{ij\setminus k}.
\eea
where  $\sigma_{ijk}$ evaluates the contribution of the two incident multilinks while $\sigma_{ij\setminus k}$,  evaluates instead the contribution due to the existence of other multilinks,  joining node $i$ and node $j$ directly or by paths of length  two  excluding node $k$. The parameter $\epsilon\in (0,1)$  tunes the relative importance between these two contributions.
The term $\sigma_{ijk}$ is expressed as
\bea
\sigma_{ijk}&=&z^{\beta_{ik,jk}},\nonumber \\
\eea
with
\bea
\beta_{ij,rs}=1-\frac{\sum_{\alpha=1}^M{m}_{ij}^{[\alpha]} {m}_{rs}^{[\alpha]}}{M}.
\label{alpha}
\eea
The smaller is $z$ the larger is the ``penalty'' for having  multilinks $\vec{m}_{ik}$ and $\vec{m}_{jk}$ with different layer composition.  If the multilinks connecting nodes $(i,k)$ and $(j,k)$ have  not even a link in a common layer,  $\beta_{ik,jk}=1$ and $z^{\beta_{ik,jk}}=z$, indicating the maximum cost attributed to multiplexity. If, on the contrary the two multilinks have the same layer composition, then $\beta_{ik,jk}=0$ and $z^{\beta_{ik,jk}}=1$ indicating that we attribute  no cost penalty to this configuration.

The term $\sigma_{ij\setminus k}$ includes  contributions from paths of length one (${\cal M}_{ij}$) and two ($\hat{\cal M}_{ijr}$) between node $i$ and node $j$ that pass through node $r$ with $r\neq k$, i.e.
\bea
\sigma_{ij\setminus k}&=&\frac{1}{\mu}\left[{\cal M}_{ij}+\sum_{r\neq k}\hat{\cal M}_{ijr}\right],
\eea
where $\mu$ is a normalization constant with $\mu=\max(1,\nu)$ with 
\bea
\nu=\min\left(\sum_{r\neq k}A_{ir},\sum_{r\neq k}{A}_{jr}\right).
\eea

Similarly to the modularity measure \cite{newman2006modularity},  term ${\cal M}_{ij}$ evaluates the significance of the observed multilink $\vec{m}_{ij}$ against its expectation and, $\hat{\cal M}_{ijr}$ evaluates the significance of two non-trivial multilinks $\vec{m}_{ir},\vec{m}_{jr}$ connecting respectively node $i$ and node $j$ to a common node $r\neq k$ against their expectations. These terms are 
\bea
{\cal M}_{ij}&=&(A_{ij}-p_{ij}^{\vec{m}_{ij}})z^{\beta_{ij,ij}}\delta\left({A}_{ij},1\right),\nonumber \\
\hat{\cal M}_{ijr}&=&(A_{ir}A_{jr}-p_{ir}^{\vec{m}_{ir}}p_{jr}^{\vec{m}_{jr}})z^{\beta_{ir,jr}}\delta\left(A_{ir}A_{jr},1\right),
\eea
where $\beta_{ij,rs}$ is given by Eq. $(\ref{alpha})$, and $\delta(x,y)$ is the Kronecker delta (i.e. $\delta(x,y)=1$ for $x=y$ and $\delta(x,y)=0$ otherwise). The term $z^{\beta_{ij,rs}}$ puts a cost to the paths that are created using different layers. The expectation of multilink $\vec{m}_{rs}$ is given by the probability $p_{rs}^{\vec{m}_{rs}}$, which is evaluated using  maximum entropy ensembles preserving the degree of node $i$ and node $j$ in each layer $\alpha$, and the multilinks $\vec{m}_{ik}, \vec{m}_{jk}$ (see Methods and SI for details).

 \subsection*{Multilink Communities} From the $L \times L$  similarity matrix $S_{ik,js}$, we construct  a dendrogram via single linkage hierarchical clustering. 
The dedrogram contains information about the multiplex structure which cannot be obtained from the aggregated network. Finally the multilink communities  are determined by cutting  the dendrogram at a height that correspond to an optimal value of a appropriate score function. 
 
To obtain the multilink communities we desire to use a score function that does not use any a priori assumptions about the multilink composition. To this end we have considered a score function used on single-layer link-community detection methods, i.e. the link modularity ${\mathcal Q}$ \cite{evans2009line} (see Method for its definition). An alternative choice could be to choose the partition density $D$ used in \cite{ahn2010link}. The optimal partition is defined by the maximum value of ${\mathcal Q}$ obtained when considering all the heights in the dendrogram (see SI for typical profiles of this link modularity on real datasets).

Once every multilink is associated to a given multilink community we can assign to each node a {\em community activity} given by the number of communities to which its incident multilinks belong.
\begin{figure}[!htb]
\begin{center}
\includegraphics[width=5.4cm]{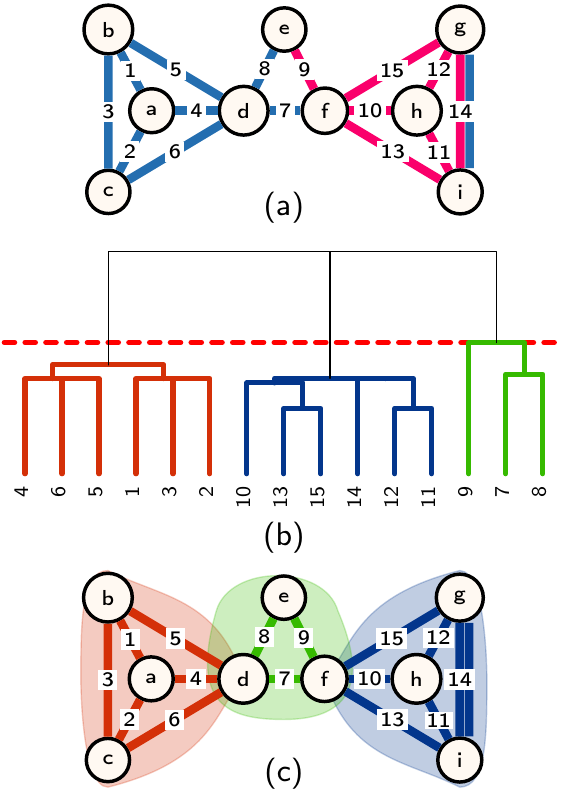}
\end{center}
\caption{(a) A simple multiplex and (b) its dendrogram obtained from the multilink similarity. The dashed red line shows the maximum link modularity used to define the link communities. Panel (c) shows  the partition of the multiplex network into three communities revealing that communities can be formed by a single (community $\{a,b,c,d\}$) or multiple layers (community $\{d,e,f\}$) and that the nodes communities are independent on the node activity (node $d$ belongs to two community and is active in one layer, node $g$ is belongs to one community and is active in one layer). The multilink communities are detected using $\epsilon=0.4,z=0.6$.}
\label{fig:benchmark}
\end{figure}
\section*{DISCUSSION}

\subsection*{A simple example} The community activity of a node resulting from the multilink community detection method  is independent on its layer activity.
To illustrate this property we consider the multilayer network shown in Fig.~\ref{fig:benchmark}(a) decomposed in  three multilink communities \ref{fig:benchmark}(c)) detected using the parameters $\epsilon=0.4$ and $z=0.6$. 
Node $d$ is active in a single layer but belongs to two multilink communities. 
On the contrary node $g$ is active in two layers but belongs to just one community.

Additionally the communities can be formed by interactions existing only in one layer or in multiple layers. For instance the  community formed by the nodes $\{e,d,f\}$ of the multiplex network shown in Fig.~\ref{fig:benchmark}(a),  only exist due to the combination of different layers in the multiplex. On the contrary the community formed by the nodes $\{a,b,c,d\}$ include only links of a single layer.

The  dendrogram in Fig.~\ref{fig:benchmark}(b) shows the hierarchical structure of the link communities of the multiplex network in Fig.\ref{fig:benchmark}(a)  and reveals the multilayer nature of the network also in the case of this very symmetrical and clustered topology.
 In fact, the left and right communities of Fig.~\ref{fig:benchmark}(c), although  they  play the same role in the aggregated network,  have a different decomposition into multilink sub-communities. There are two factors that contribute to this difference. The right community has a multilink formed by two layers (multilink 14) which is no present in the other community. The second factor is more subtle and it would generate differences in the hierarchical structure even if the community on the right included only links existing in a single layer (see SI for details). 
 
\begin{figure}[!htb]
\begin{center}
\includegraphics[width=7.1cm]{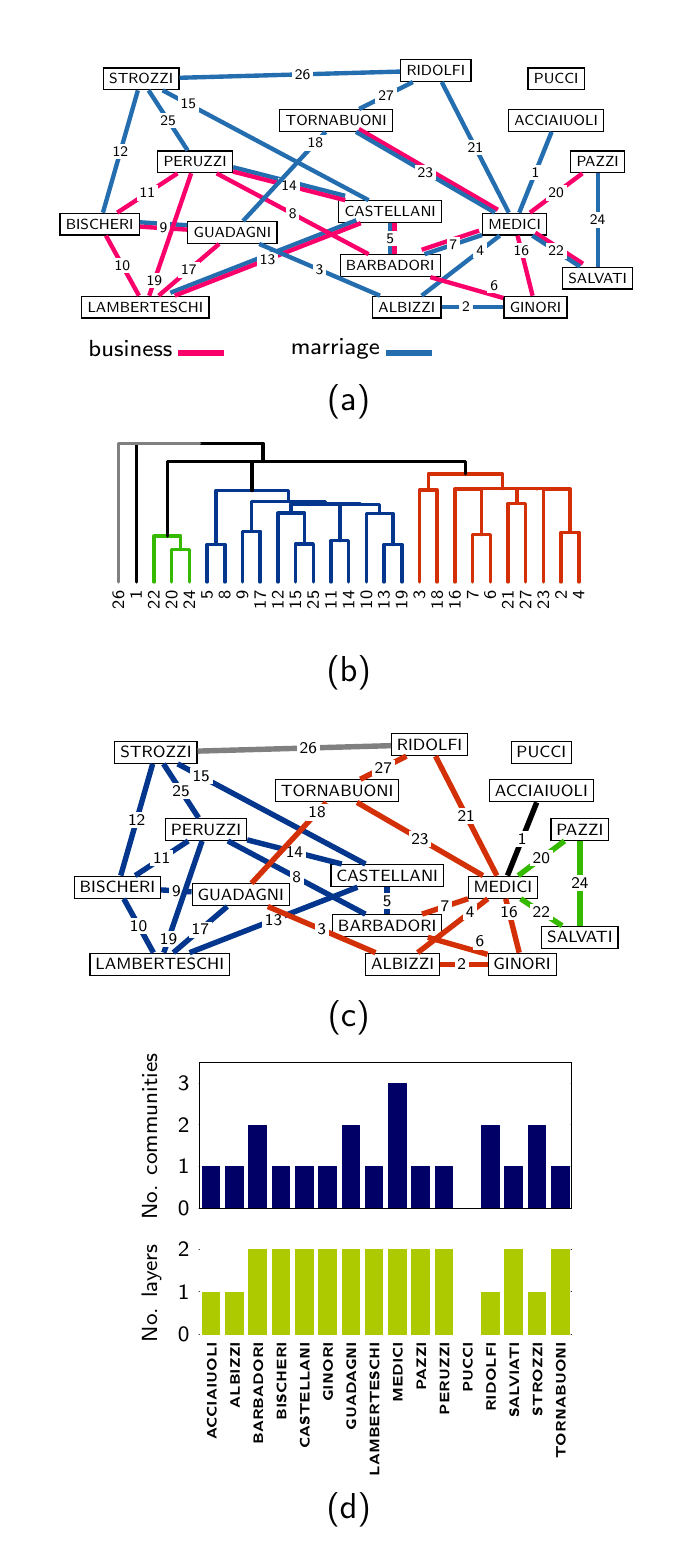}
\end{center}
\caption{(a) The Florentine Families Multiplex Nework  describing the business and marriage alliances of the XV century florentine families. (b) Heat map displaying the multilink similarity matrix and its relative dendrogram . (c) Partition of the Florentine Families multiplex network into five multilink communities. (d) Layer and community activity of the different families. The Medici family is characterized by achieving  the maximum of the  community activity. The multilink communities are detected using $\epsilon=0.4,z=0.6$.}
\label{fig:florentine}
\end{figure}
\subsection*{Florentine Families} The Florentine Families Multiplex Network  \cite{padgett1993robust} consist of $M=2$ two layers, one layer describes the business dealings between $N=16$ florentine families in the {XV century}, the other layer their alliances due to marriages. 
Fig.~\ref{fig:florentine}(a) shows these relationships between the families. 
Figure~\ref{fig:florentine}(b) shows the dendrogram describing the multilink communities for $\epsilon=0.5, z=0.6$ (see SI for the dependence of the number of clusters on $\epsilon$ and $z$). 

The  two detected single multilink communities  correspond to two different scenarios (Fig.~\ref{fig:florentine}(c)). The multilink between the Strozzi and the Ridolfi family establish an interaction between two families which have connections between different clusters; the multilink between the Acciaiuoli and the Medici family is a leaf of the multiplex network, being the only multilink connecting the Acciaiuoli family to the rest of the multiplex network. 

For each family we compare their {\em layer activity}  and their {\em community activity} (Fig.~\ref{fig:florentine}(d)). We observe that families with high community activity are powerful brokers between different communities.
Most relevantly, the Medici play a pivotal role as they are brokers between three different communities. 
The Barbadori and the Guadagni family have the same community activity as the Ridolfi and the Strozzi family but while the first two are connected in both layers the latter two are connected to the other families exclusively in one layer (the marriage alliances).

\begin{figure}[!htb]
\begin{center}
\includegraphics[width=8.2cm]{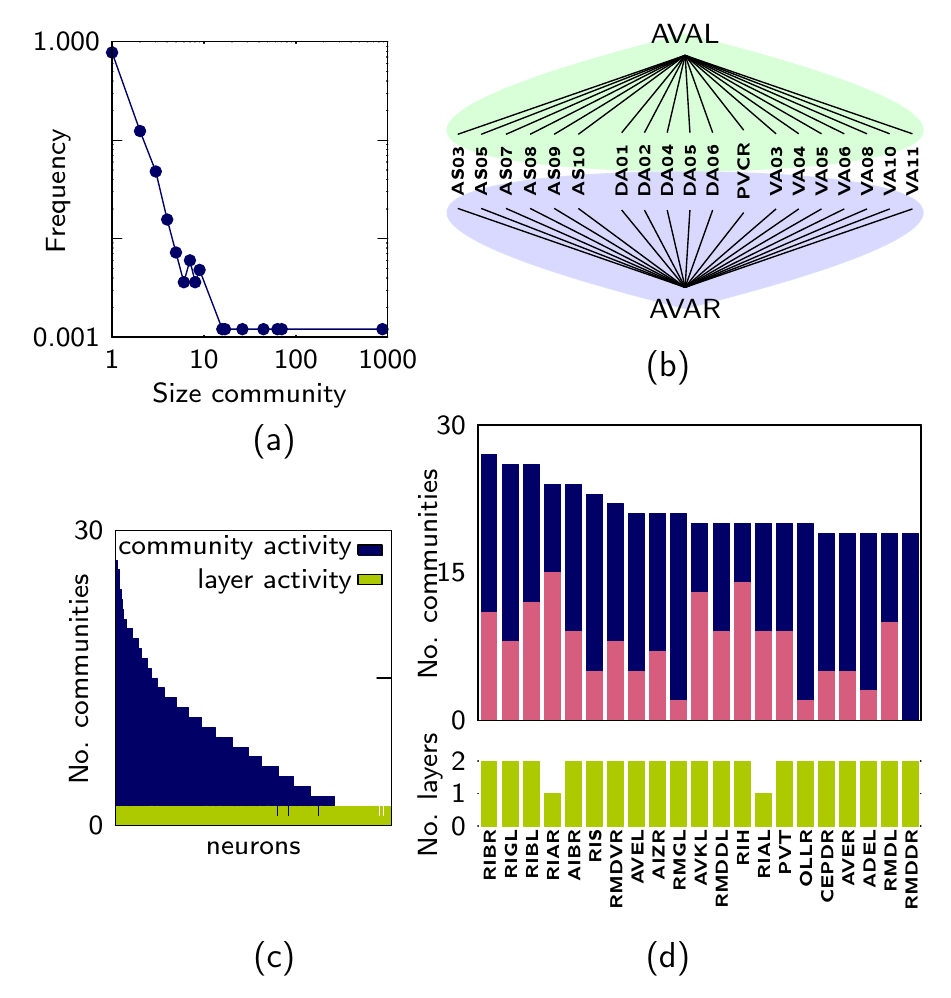}
\end{center}
\caption{(a) Distribution of the communities sizes for the Multiplex Connectome of  {\em C. elegans}. (b) The two most similar sub-communities contained in the largest multilink community. (c) Neurones ranked in decreasing order of their community activity. (d) Layer and community activity for the top ranked neurons. (the contribution of communities with single multilinks is in pink while the contributions of communities with more than one multilink is in blue). The multilink communities are detected using $\epsilon=0.4,z=0.6$.}
\label{fig:celegans}
\end{figure}
\subsection*{Multiplex Connectome of C. elegans}  The Multiplex Connectome of {\it C.~elegans} \cite{chen2006wiring,de2014muxviz} has two layers $M=2$, the chemical synapses and the gap junctions describing the interactions between $N=279$ neurons. 
As an example, we obtained the multilink communities for $\epsilon=0.4$ and $z=0.6$. The multiplex has 845 multilink communities of which 652 (about 77$\%$) are made of single multilinks. The distribution of the sizes of the communities is broad. (Fig.~\ref{fig:celegans}(a)). The largest community is formed by 878 multilinks followed communities including 67 links and 51 links.
Although there is a large dominant community in the multiplex network, the internal structure of this community can be investigated via the dendrogram. We noticed that the ADAL and ADAR are the neurones that cluster first with some of their neighbouring neurones (Fig.~\ref{fig:celegans}(b)) for all values of $z$.

This multiplex has neurons which have large community activity (Fig.~\ref{fig:celegans}(c)). By ranking the neurons according to their community activity we find in the first two positions the RIBR and RIBL neurons, which are head interneurons connected via gap junctions to multiple other neuron classes, suggesting that these neurons play a role in brokering between different communities (Fig.~ \ref{fig:celegans}(d)).
\begin{figure}[!htb]
\begin{center}
\includegraphics[width=8.3cm]{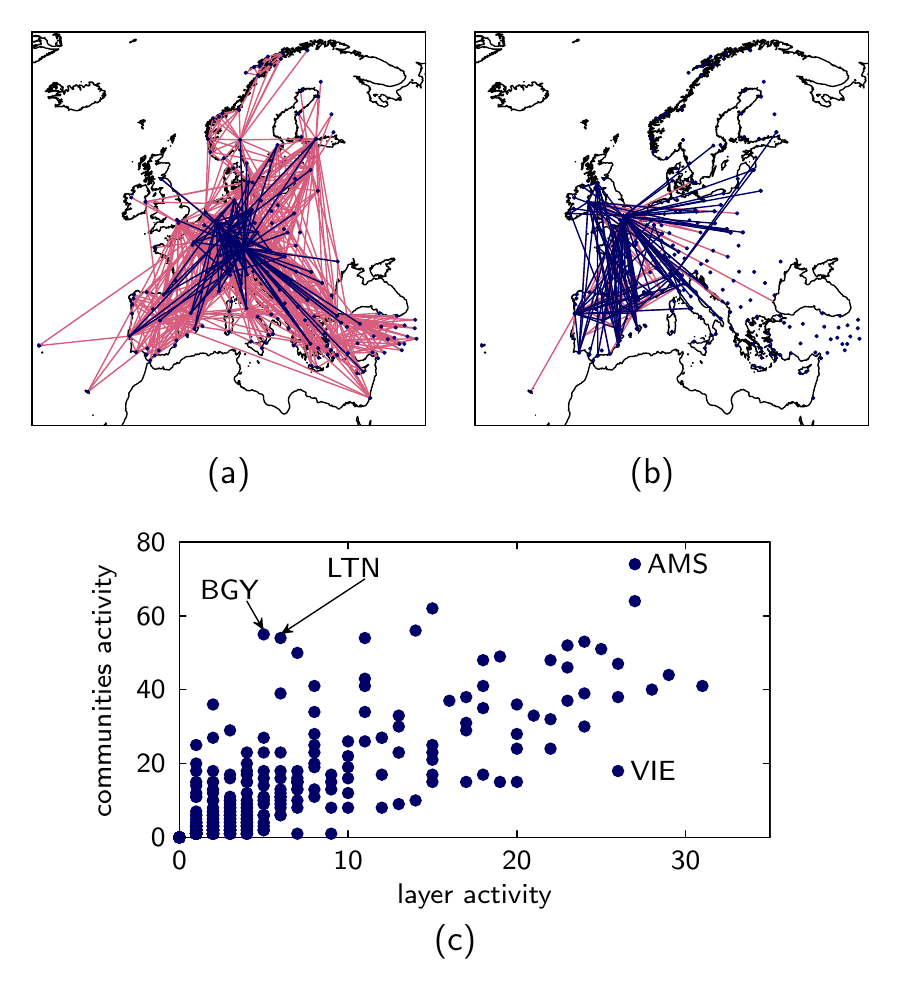}
\end{center}
\caption{(a) Largest link community of the European Multiplex Air Transportation Network. Lufthansa's flights are shown in blue, the other airlines in pink. (b) Second largest link community. Ryanair's flights are shown in blue, the other airlines in pink. (c) Community vs. layer activity of the EU airports. While the layer activity appears to have a positive correlation with the layer activity, large difference in community activity can be oserved between airports with large layer activity (compare for instance Amsterdam (AMS) and Vienna (VIE)).The multilink communities are detected using $\epsilon=0.4,z=0.6$.}
\label{fig:airports}
\end{figure}
\subsection*{European Multiplex Air Transport Network} The European Multiplex Air Transport Network~\cite{cardillo2013emergence} comprises of $N=417$ European airports and $M=37$ layers corresponding to the airlines that have flight connections between these airports. The total number of multilink describing these connections is 2953.
For the case that $\epsilon=0.4$ and $z=0.6$, our algorithm obtains 1790 multilink communities. The largest community includes 723 nodes, about 24\% of the total number of multilinks. The smallest communities are made of single multilinks and there are 1696 of them, about 57\% of the multilinks.

We observe that the main communities have very different composition in term of single layers. Figure~\ref{fig:airports}(a)-(b) shows the two largest communities.   All the airlines (layers) contribute to the structure of the largest community (Fig.~\ref{fig:airports}(a)).  The second largest community has a very different structure, only few airlines contribute to this community.

When comparing the airports and their community activity, we observe (Fig.~\ref{fig:airports}(c)) that while large layer activity, an airport serving multiple airline companies, seems to be correlated to high community activity, there is a significant variability in the community of airports that are active in many layers. For example Vienna (VIE) and Amsterdam (AMS) have a comparable layer activity but very different community activity. Similarly there are airports with small layer activity but significant community activity, for example Luton (LTN) and Bergamo (BGY) airports. 
This indicates that the airports might adopt different strategies to  broker between different communities. These strategies might involve   serving flights of many airline companies or serving flights of relatively fewer airline companies.

\begin{figure}[!htb]
\begin{center}
\includegraphics[width=8.6cm]{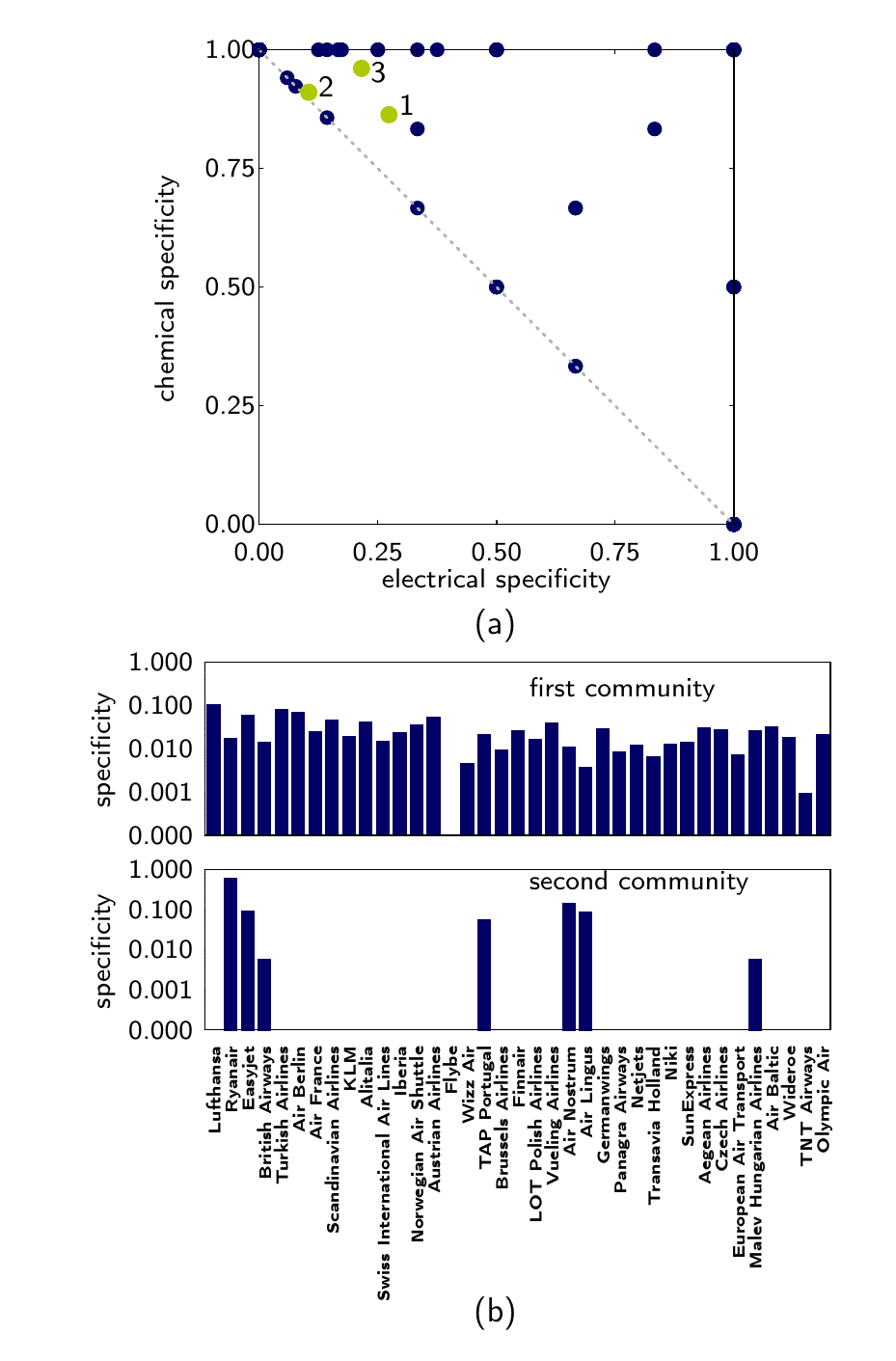}
\end{center}
\caption{(a) Specificity for the communities in the Multiplex Connectome of {\em C. elegans}. (b) Specificity for the first and second largest communities for the European Air Transportation Network. Both panels indicate a large variability in the layer composition of different communities. The multilink communities are detected using $\epsilon=0.4,z=0.6$.}
\label{fig:specificity}
\end{figure}
\subsection*{Composition of the multilink communities} To investigate whether the communities are formed exclusively by links of a single layer or  include links of several layers we introduce the 
 {\em layer specificity} $x^{[\alpha]}_c$ which is the fraction of multilinks in a community $c$ which include a link in layer $\alpha$.
Therefore $x^{[\alpha]}_c=1$ indicates that all the multilinks of a community include a link in layer $\alpha$, while $x^{[\alpha]}_c=0$ indicates that the community does not include any link in layer $\alpha$. 
Note that since a single multilink can include links of different layers, the sum of the layer specificity $x^{[\alpha]}_c$ for community $c$ in general do not add to one.

In the  Multiplex Connectome of {\sl C.~elegans}  we observe that many communities are exclusively formed by one type of multilink, however, the three largest communities have a multiplex nature as they include different types of multilinks (see Fig. $\ref{fig:specificity}(a)$ where the larger communities are indicated by the labels 1, 2, 3 in order of decreasing size).

In the  European Multiplex Air Transportation Network, the largest community, apart from Flybe, contains flights from all other airlines (Fig~\ref{fig:specificity}(b)). The largest contribution comes from Lufthansa with an specificity of 0.10 followed by Turkish Airlines with 0.07 specificity. The second largest community has a different structure, in this case only seven airlines contribute to the community, the largest contribution is from Ryanair with a specificity of 0.60.
 In this multiplex, low-cost airlines like Ryanair, Easyjet and Wideroe have high specificity (often equal to 1) in many communities. However these airlines rarely have high specificity in the same community. This is a consequence of the competition between low-cost airline companies as they tend to differentiate each other by having unique flights to some destinations.

\section*{CONCLUSIONS}
Our method reveal the richness of multiplex networks at their mesoscale structure. This is achieved by associating to each pair of incident multilinks a similarity measure based on the comparison of  the local connectivity of two  multilinks against a null model.  Our intrinsically multiplex community detection method  allow us to associate to each node multiple communities independently on its layer activity. Specifically we can have nodes active exclusively in one layer and belonging to multiple communities or active in many layers but belonging only to few communities. The proposed method is here applied to several real datasets revealing that the  mesoscale structure of a multiplex can be organised via communities containing links in many different layers and, at the same time, communities having one predominat layer. This suggests that the mesoscale organization of multiplex networks has a rich mesoscale structure that is not captured by methods that aim at  compressing  the information on few single layers.
\section*{MATERIALS AND METHODS}{
\subsection*{Maximum Entropy Ensemble}
To evaluate  the similarity $S_{ik;jk}$ between two incident multilinks connecting nodes $(i,k)$ and $(j,k)$ we need to calculate the probability $p_{\ell r}^{\vec{m}_{\ell r}}$ of the multilinks $\vec{m}_{\ell r}$ with $\ell=i,j$ and $r\neq k$ in our null model. The null model is a maximum entropy ensemble determined by the 
 probability $P(\vec{\mathcal {G}})$ associated to each possible  multiplex network $\vec{\mathcal {G}}$ with  adjacency matrices ${\bf \tilde{a}^{[\alpha]}}$ and satisfying the  constraints
\bea
\sum_{\vec{\mathcal{G}}}\left(P(\vec{\mathcal{G}}) \sum_{r \neq k } \tilde{a}_{\ell r}^{[\alpha]}\right)&=&q_\ell^{[\alpha]}-a_{\ell k}^{[\alpha]},
\eea
with $\alpha=1,2,\ldots, M$, $\ell=i,j$ and $q_{\ell}^{[\alpha]}$ indicating the degree on node $\ell$ in layer $\alpha$.(See SI for further details).
Therefore the ensemble randomizes the original multiplex network by keeping constants the degrees $q_{\ell}^{[\alpha]}$ and the multilinks $\vec{m}_{\ell k}$ with $\ell=i,j$.
 
\subsection*{Link modularity}

Let us  consider the adjacency matrix ${\mathbf W}$  determining the line graph of the aggregated network. This matrix has elements $W_{\ell,\ell'}=1$ if the link $\ell$ is incident to the link $\ell'$ while otherwise $W_{\ell,\ell'}=0$. 
For any given dendrogram cut, we indicate the cluster membership of multilink corresponding to the link $\ell$ of the aggregated network as $c_{\ell}$.
The link modularity $\mathcal{Q}$ \cite{evans2009line} is given by 
 \bea
 \mathcal{Q}=\frac{1}{\sum_{\ell} d_{\ell}}\sum_{\ell,\ell'}\left[W_{\ell,\ell'}-\frac{d_{\ell}d_{\ell'}}{\sum_{\ell}d_{\ell}}\right]\delta(c_{\ell},c_{\ell'}),
 \eea
 where $d_{\ell}=\sum_{\ell'}W_{\ell,\ell'}$ and $\delta(a,b)=1$ if and only if $a=b$ otherwise $\delta(a,b)=0$.

\subsection*{Codes}
The codes implementing the Multilink Community detection method are freely available at the website: https://github.com/ginestrab.
\subsection*{Data}
All the datasets analyzed in this paper are freely available on the  data repository http://deim.urv.cat/\textasciitilde manlio.dedomenico/data.php.
}

\section*{ACKNOWLEDGMENTS}
This research utilised Queen Mary's MidPlus computational facilities, supported by QMUL Research-IT and funded by EPSRC grant EP/K000128/1. 

\newpage
\newpage

\section*{{ SUPPLEMENTARY INFORMATION}}

\renewcommand\theequation{{S-\arabic{equation}}}
\renewcommand\thetable{{S-\Roman{table}}}
\renewcommand\thefigure{{S-\arabic{figure}}}

\section*{Supplementary Information on the Multilink Community detection algorithm}
\subsection*{General considerations}

The similarity between any two incident multilinks of the multiplex network is the basic element of the multilink community detection algorithm. The similarity matrix is used  to perform a hierarchical clustering of the multilinks, ultimately finding the multilink communities as described in the main body of the paper. In the same spirit as in Ref. \cite{newman2006modularity} the  similarity matrix is constructed by comparing the local neighborhood of each pair of incident multilinks to a maximum entropy null model for the multiplex network.

Here we give further information on the maximum entropy null model that we used to evaluate the similarity between any two incident multilinks. This model extends previous results on exponential random graphs of single \cite{park2004statistical}   and multiplex networks \cite{bianconi2013statistical,menichetti2014weighted}. 
\subsection*{Multiplex network}
Let us consider a multiplex network $\vec{G}=\{G^{[1]},G^{[2]},\ldots, G^{[\alpha]},\ldots, G^{[M]}\}$ formed by $N$ nodes and $M$ layers $\alpha=1,2,\ldots, M$. . Every layer $\alpha$ is formed by a  undirected network with  adjacency matrix ${\bf a}^{[\alpha]}$.
Every pair of nodes $(i, j)$ is connected by a multilink~\cite{bianconi2013statistical,menichetti2014weighted} 
\bea
\vec{m}_{ij}=\left(m_{ij}^{[1]},m_{ij}^{[2]},\ldots,m^{[\alpha]}_{ij}\ldots m_{ij}^{[M]}\right),
\eea
with $m_{ij}^{[\alpha]}=a_{ij}^{[\alpha]}$ indicating in which layers of the multiplex network the two nodes are connected. Whenever node $i$ and node $j$ are connected at least in one layer, i.e. $\vec{m} \neq\vec{0},$ we say that they are connected by a non-trivial multilink. 

{The aggregated network $\hat{G}$  is the single network in which any two nodes are connected if they are linked at least in one layer of the multiplex network. The adjacency matrix ${\bf A}$ of the aggregated network $\hat{G}$ has elements }
\bea
{A}_{ij}=\theta\left(\sum_{\alpha=1}^M a_{ij}^{[\alpha]}\right),
\eea
where $\theta(x)$ is the step function $\theta(x)=1$ if $x>0$ and $\theta(x)=0$ if $x\leq 0$. We indicate with $L=\sum_{i<j}A_{ij}$ the total number of links of the aggregated network, or equivalently the number of non-trivial multilinks.

\subsection*{Multilink similarity}
In order to detect the multilink communities we assign a non zero similarity $S_{ik,jk}$ to every pair of incident multilinks connecting respectively the generic nodes $i$ and $k$ and $j$ and $k$. The non-zero similarities $S_{ik,jk}$ are given by
\bea
S_{ik,jk}=\epsilon \sigma_{ijk}+(1-\epsilon)\sigma_{ij\setminus k}.
\eea
where  $\sigma_{ijk}$ evaluates the contribution of the two incident multilinks while $\sigma_{ij\setminus k}$,  evaluates instead the contribution due to the existence of other multilinks,  joining node $i$ and node $j$ directly or by paths of length  two  excluding node $k$. The parameter $\epsilon\in (0,1)$  tunes the relative importance between these two contributions.
The term $\sigma_{ijk}$ is expressed as
\bea
\sigma_{ijk}&=&z^{\beta_{ik,jk}},\nonumber \\
\eea
with
\bea
\beta_{ij,rs}=1-\frac{\sum_{\alpha=1}^M{m}_{ij}^{[\alpha]} {m}_{rs}^{[\alpha]}}{M}.
\label{alpha}
\eea
The term $\sigma_{ij\setminus k}$ includes  contributions from paths of length one (${\cal M}_{ij}$) and two ($\hat{\cal M}_{ijr}$) between node $i$ and node $j$ that pass through node $r$ with $r\neq k$, i.e.
\bea
\sigma_{ij\setminus k}&=&\frac{1}{\mu}\left[{\cal M}_{ij}+\sum_{r\neq k}\hat{\cal M}_{ijr}\right],
\eea
where $\mu$ is a normalization constant with $\mu=\max(1,\nu)$ with 
\bea
\nu=\min\left(\sum_{r\neq k}A_{ir},\sum_{r\neq k}{A}_{jr}\right).
\eea

Similarly to the modularity measure \cite{newman2006modularity},  term ${\cal M}_{ij}$ evaluates the significance of the observed multilink $\vec{m}_{ij}$ against its expectation and, $\hat{\cal M}_{ijr}$ evaluates the significance of two non-trivial multilinks $\vec{m}_{ir},\vec{m}_{jr}$ connecting respectively node $i$ and node $j$ to a common node $r\neq k$ against their expectations. These terms are 
\bea
{\cal M}_{ij}&=&(A_{ij}-p_{ij}^{\vec{m}_{ij}})z^{\beta_{ij,ij}}\delta\left({A}_{ij},1\right),\nonumber \\
\hat{\cal M}_{ijr}&=&(A_{ir}A_{jr}-p_{ir}^{\vec{m}_{ir}}p_{jr}^{\vec{m}_{jr}})z^{\beta_{ir,jr}}\delta\left(A_{ir}A_{jr},1\right),
\eea
where $\beta_{ij,rs}$ is given by Eq. $(\ref{alpha})$, and $\delta(x,y)$ is the Kronecker delta (i.e. $\delta(x,y)=1$ for $x=y$ and $\delta(x,y)=0$ otherwise). The expectation of multilink $\vec{m}_{rs}$ is given by the probability $p_{rs}^{\vec{m}_{rs}}$, which is evaluated using  maximum entropy ensembles.

 The null model should not change the multilinks $\vec{m}_{ik}$ and $\vec{m}_{jk}$ determining the connection of nodes $i$ and $j$ with node $k$. This restriction fixes the connections between node $i$ and $k$ and node $j$ and $k$ but it does not restrict the connections between nodes $i$ and $j$ and their other neighbors. To capture the local structure on layer $\alpha$, the null model should preserve the number of neighbors of nodes $i$ and $j$ in each layer $\alpha$, that is their degree $q^{[\alpha]}_{i}$ and $q^{[\alpha]}_{j}$, however, except from node $k$, the neighbors are selected at random from the remaining $N-2$ nodes. 
Therefore the maximum entropy model is preserving the degree of node $i$ and node $j$ in each layer $\alpha$, and the multilinks $\vec{m}_{ik}, \vec{m}_{jk}$.

\subsection*{Maximum entropy ensemble}
The considered maximum entropy  ensemble is characterised by the probability $P(\vec{\mathcal{G}})$ assigned to each possible multiplex network $\vec{\mathcal{G}}$ determined by the set of adjacency matrices ${\bf \tilde{a}}^{[\alpha]}$ with $\alpha=1,2,\ldots, M$.
This probability is found by maximising the entropy $S$ which is the logarithm of the number of typical multiplex networks in the ensemble, 
\bea
S=-\sum_{\vec{\mathcal{G}}}P(\vec{\mathcal{G}})\ln P(\vec{\mathcal{G}})
\eea
given the set of structural constraints under consideration.
These constraints are
\bea
\sum_{\vec{\mathcal{G}}}\left(P(\vec{\mathcal{G}}) \sum_{r \neq k } \tilde{a}_{ir}^{[\alpha]}\right)&=&q_i^{[\alpha]}-a_{ik}^{[\alpha]},\nonumber \\
\sum_{\vec{\mathcal{G}}}\left(P(\vec{\mathcal{G}}) \sum_{r \neq k } \tilde{a}_{jr}^{[\alpha]}\right)&=&q_j^{[\alpha]}-a_{jk}^{[\alpha]},
\label{const}
\eea
with $\alpha=1,2,\ldots, M$.
By introducing the Lagrangian multipliers $\lambda_i^{[\alpha]}, \lambda_j^{[\alpha]}$ with $\alpha=1,2,\ldots, M$ the probability $P(\vec{\mathcal{G}})$ can be written as
\bea
P(\vec{\mathcal{G}}) =\frac{1}{Z}e^{-\sum_{\alpha=1}^M H_{ij}^{[\alpha]}},
\eea
where the partition function $Z$ is a normalization constant, and $F_{ij}^{[\alpha]}$ is given by 
\bea
H_{ij}^{[\alpha]}&=&\lambda_i^{[\alpha]}\left(\sum_{r \neq \{k,i,j\}}\tilde{a}_{ir}^{[\alpha]}\right)+\lambda_j^{[\alpha]}\left(\sum_{r \neq \{k,i,j\}}\tilde{a}_{jr}^{[\alpha]}\right)\nonumber \\
&&+(\lambda_i^{[\alpha]}+\lambda_j^{[\alpha]})\tilde{a}_{ij}^{[\alpha]}.
\eea

The marginal probability of single links of nodes $i$ and node $j$ in each layer $\alpha$ are given, for $r\neq \{i, j, k\}$ by
\bea
p_{ir}^{[\alpha]}&=&\sum_{\vec{\mathcal{G}}}\left(P(\vec{\mathcal{G}})\tilde{a}_{ir}\right)=\frac{e^{-\lambda_i^{[\alpha]}}}{1+e^{-\lambda_i^{[\alpha]}}},
\nonumber \\
p_{jr}^{[\alpha]}&=&\sum_{\vec{\mathcal{G}}}\left(P(\vec{\mathcal{G}})\tilde{a}_{jr}\right)=\frac{e^{-\lambda_j^{[\alpha]}}}{1+e^{-\lambda_j^{[\alpha]}}},
\eea
and by
\bea
p_{ij}^{[\alpha]}&=&\sum_{\vec{\mathcal{G}}}\left(P(\vec{\mathcal{G}})\tilde{a}_{ij}\right)=\frac{e^{-\lambda_i^{[\alpha]}-\lambda_j^{[\alpha]}}}{1+e^{-\lambda_i^{[\alpha]}-\lambda_j^{[\alpha]}}}.
\eea
The Lagrangian multipliers $\lambda_i^{[\alpha]}$ and $\lambda_j^{[\alpha]}$ are determined by the constraints in Eq. (\ref{const}) that, in terms of the marginals is
\bea
\left(\sum_{r\neq\{ j,k\}}p_{ir}^{[\alpha]}\right)+p_{ij}^{[\alpha]}=q_i^{[\alpha]}-a_{ik}^{[\alpha]},
\nonumber \\
\left(\sum_{r\neq \{i,k\}}p_{jr}^{[\alpha]}\right)+p_{ij}^{[\alpha]}=q_j^{[\alpha]}-a_{jk}^{[\alpha]}.
\eea
Finally this maximum entropy ensemble allow us to determine the probability $p_{ir}^{\vec{m}_{ir}}$ and $p_{jr}^{\vec{m}_{jr}}$ of the multilinks $\vec{m}_{ir}, \vec{m}_{jr}$ which are given respectively by 
\bea
p_{ir}^{\vec{m}_{ir}}&=&\sum_{\vec{\mathcal{G}}}P(\vec{\mathcal{G}})\prod_{\alpha=1}^{M}\left(\tilde{a}_{ir}^{[\alpha]}m_{ir}^{[\alpha]}+(1-\tilde{a}_{ir}^{[\alpha]})(1-p_{ir}^{[\alpha]})\right)\nonumber \\
&=&\prod_{\alpha=1}^{M}\left(p_{ir}^{[\alpha]}m_{ir}^{[\alpha]}+(1-m_{ir}^{[\alpha]})(1-p_{ir}^{[\alpha]})\right),
\eea
and
\bea
p_{jr}^{\vec{m}_{jr}}&=&\sum_{\vec{\mathcal{G}}}P(\vec{\mathcal{G}})\prod_{\alpha=1}^{M}\left(\tilde{a}_{jr}^{[\alpha]}m_{jr}^{[\alpha]}+(1-\tilde{a}_{jr}^{[\alpha]})(1-p_{jr}^{[\alpha]})\right)\nonumber \\
&=&\prod_{\alpha=1}^{M}\left(p_{jr}^{[\alpha]}m_{jr}^{[\alpha]}+(1-m_{jr}^{[\alpha]})(1-p_{jr}^{[\alpha]})\right).
\eea

\subsection*{Multilink communities}
From the $L \times L$  similarity matrix $S_{ik,js}$, we construct  a dendrogram via single linkage hierarchical clustering. 
The multilink communities  are obtained by cutting  the dendrogram at a height that correspond to the maximum value of the link modularity $\mathcal{Q}$.

The link modularity $\mathcal{Q}$ \cite{evans2009line} is given by 
 \bea
 \mathcal{Q}=\frac{1}{\sum_{\ell} d_{\ell}}\sum_{\ell,\ell'}\left[W_{\ell,\ell'}-\frac{d_{\ell}d_{\ell'}}{\sum_{\ell}d_{\ell}}\right]\delta(c_{\ell},c_{\ell'}),
 \label{Qs}
 \eea
 where ${\mathbf W}$ is the adjacency matrix of the line graph of the aggregated network and has elements $W_{\ell,\ell'}=1$ if the link $\ell$ is incident to the link $\ell'$ while otherwise $W_{\ell,\ell'}=0$. Additionally in Eq. $(\ref{Qs})$ we indicate with $d_{\ell}$ the link-degree $d_{\ell}=\sum_{\ell'}W_{\ell,\ell'}$ and   and with $c_{\ell}$  the cluster membership of the multilink corresponding to the link $\ell$ of the aggregated network. Finally $\delta(a,b)=1$ if and only if $a=b$ otherwise $\delta(a,b)=0$.

Once every multilink is associated to a given  multilink community, each node is attributed a {\em community activity} given by the number of different communities to which its incident multilinks belong.

\section*{Supplementary Information on the results obtained on real datasets with the Multilink Community detection algorithm}

\subsection*{More on the benchmark multiplex network}
\begin{figure}[!htb]
\begin{center}
\includegraphics[width=7cm]{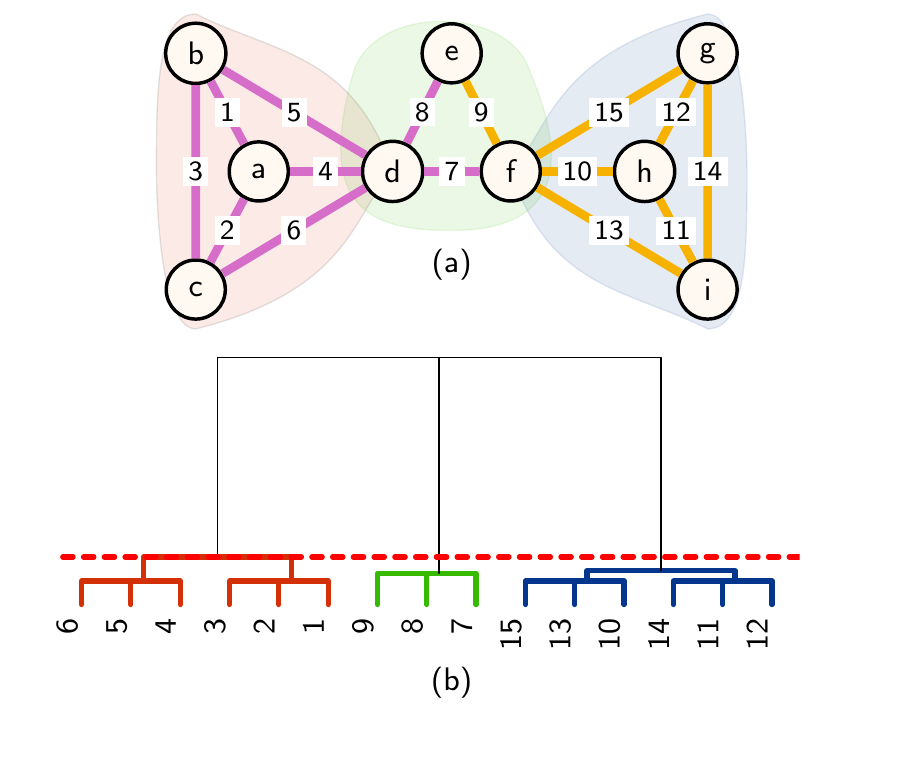}
\end{center}
\caption{(a) A simple two layer multiplex network (purple and ochre links) and its multilink communities (shaded areas)  and (b) its dendrogram obtained from the multilink similarity. The dashed red line shows the maximum link modularity used to define the link communities. 
}
\label{fig:S1}
\end{figure}

As mentioned in the main part of the paper when considering the example of the simple multiplex shown in Fig.~1(a)-(b). The right and left multilink communities have a different internal structure due to a subtle factor, 
%
this difference is clearly seen in the dendrogram Fig.~1(b). To explain this difference we consider a very simple multiplex.
Figure~\ref{fig:S1}(a) shows this simple multiplex network and its partition into multilink communities (shaded areas). Although the  community structure of this multiplex network is identical to Fig.~1(c), its dendrogram  (Fig.~\ref{fig:S1}(b)) is again,  not symmetric under the permutation of the right and left communities. The difference is due to  the multiplexity of the network. In fact node $f$ and node $d$ play slightly different roles in their communities. Node $f$ is active in  two different layers, while node $d$ is active only in one layer. Our method distinguishes these two cases.

\subsection*{Aggregated degree vs. community activity}
We investigated if there is a correlation between the degree of the aggregated network $\hat{G}$ and the community activity of the nodes. Figure~\ref{fig:S4} shows community activity vs.~degree of the aggregated network for the Multiplex Connectome of {\emph{C. elegans}} (Fig.~\ref{fig:S4}(a)) and the European Multiplex Air Transportation  Network (Fig.~\ref{fig:S4}(b)). %
For small degrees, there is a significant positive correlation between these two quantities, but as the degree increases,  the correlation diminishes.

\begin{figure}[htb]
\begin{center}
\includegraphics[width=6cm]{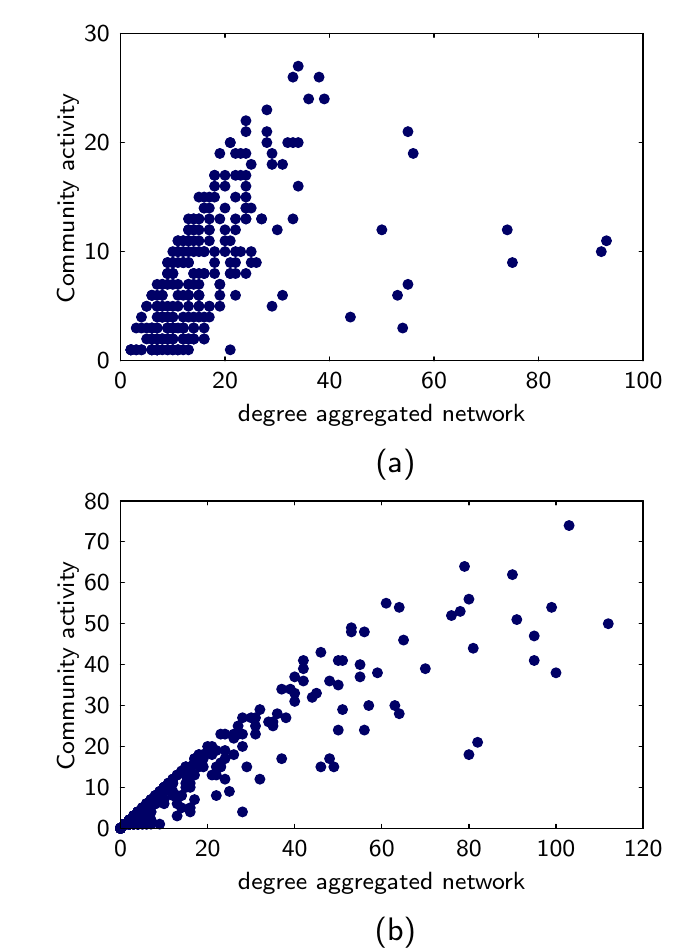}
\end{center}
\caption{Aggregated degree vs.~community activity for the (a)  Multiplex Connectome of {\emph{C. elegans}} and (b) for the  European Mutliplex Air Transportation  Network.}
\label{fig:S4}
\end{figure}

\subsection*{The score function profile of the analyzed datasets}
\begin{figure}[!htb]
\begin{center}
\includegraphics[width=6cm]{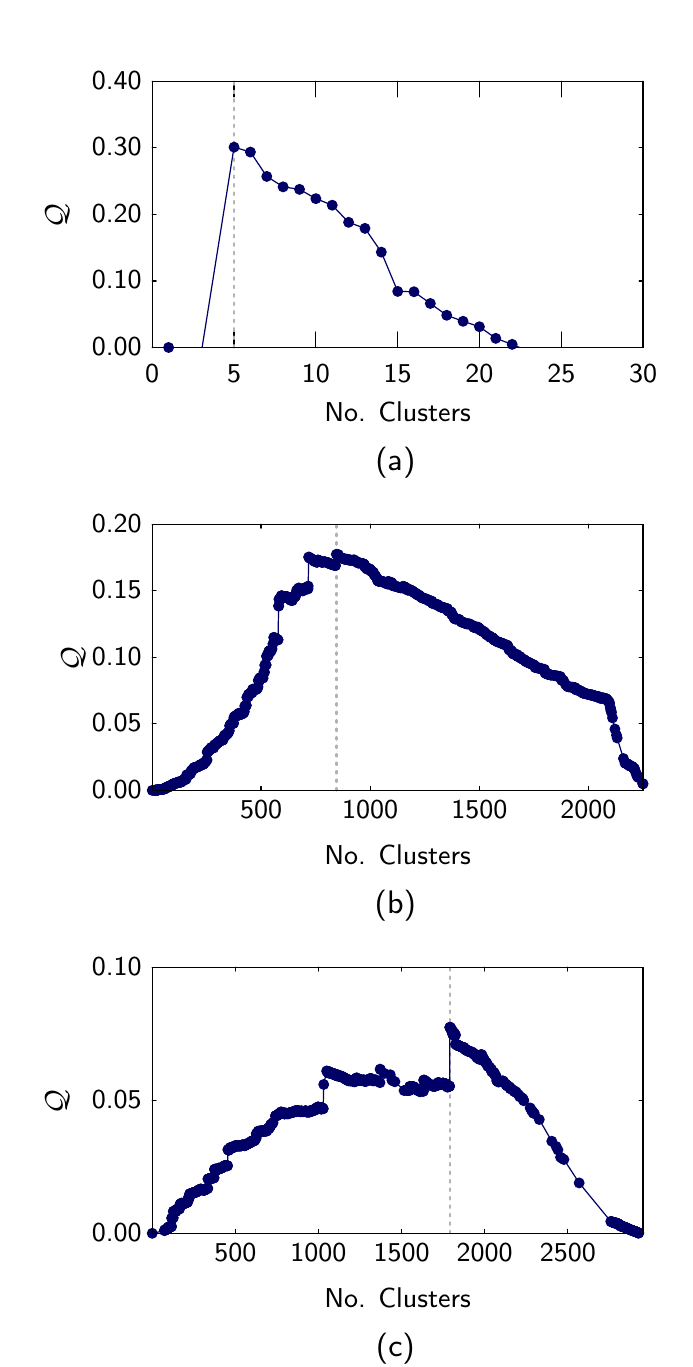}
\end{center}
\caption{Number of clusters against the score function ${\cal Q}$  (link--modularity) for the (a) Florentine Families Multiplex Network ($\epsilon=0.5, z=0.6$), (b) for the Multiplex Connectome of {\em{C. elegans}} ($\epsilon=0.4, z=0.6$) and (c) for the  European  Multiplex Air Transportation   Network ($\epsilon=0.4, z=0.6$). The maximum of ${\cal Q}$ determines the number of clusters which define the multlink communities of the multiplex network. }
\label{fig:S2}
\end{figure}

The multilink communities are determined by cutting the dendrogram at a height that corresponds to the maximum value of the score function $\mathcal{Q}$. In the datasets considered here, we observed that the profile of the link modularity $\mathcal{Q}$ (Fig.~\ref{fig:S2}) displays a  well defined global maximum,  suggesting that the determination of the optimal partition is not questionable.

\subsection*{The number of multilink communities as a function of  the parameters  $\epsilon$ and $z$}
In general the values of the parameters $z$ and $\epsilon$ will depend on the network under consideration. The parameters used in the here were chosen to demonstrate the dependence of the number of multilink communities as  function of the parameters. An example of this dependance is shown in Fig.~\ref{fig:S4}(a)-(b) for the Florentine families.  We noticed that for many different values of the parameters the method consistently divided the multiplex into five link communities, in the manuscript we used $\epsilon=0.4$ and $z=0.6$ to show these five link communities.


\begin{figure}[t]
\begin{center}
\includegraphics[width=6cm]{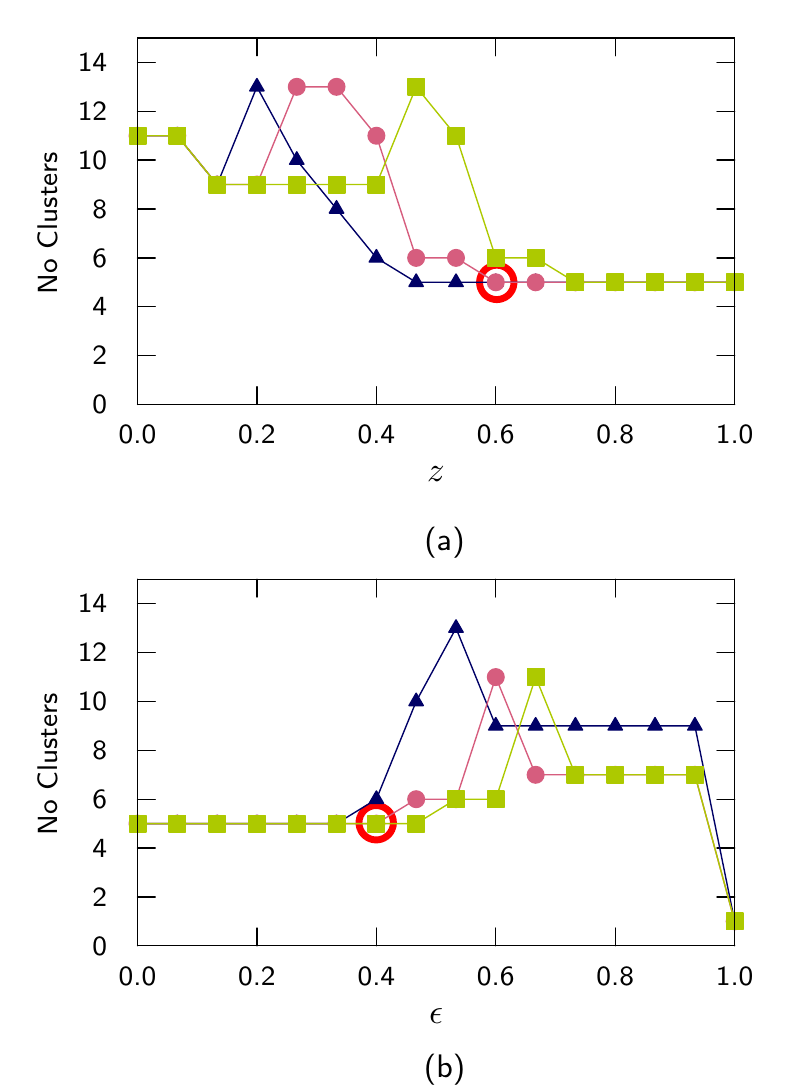}
\end{center}
\caption{(a) Variation of the number of clusters for the Florentine families multiplex as a function of $z$ where the data  shown are for $\epsilon=0.4$ (blue triangles), $\epsilon=0.5$ (pink circles) and $\epsilon=0.6$ (green squares).
(b) Variation of the number of clusters for the Florentine families multiplex as a function of $\epsilon$ where the data  shown are for  for $z=0.4$ (blue triangles), $z=0.5$ (pink circles) and $z=0.6$ (green squares).  
In both figures the red circle marks the values of the parameters used in Fig.~\ref{fig:florentine} }
\label{fig:S4}
\end{figure}

%
%
\vfill
\quad
\newpage

%
%
%

\end{document}